\documentclass{llncs}

\usepackage{amsmath,amssymb,bm}
\usepackage{url}
\usepackage{isabelle,isabellesym}
\title{Ackermann's Function in Iterative Form:\\ A Proof Assistant Experiment}
\author{Lawrence C. Paulson FRS}
\institute{Computer Laboratory, University of Cambridge, England\\ \email{lp15@cam.ac.uk}}

\DeclareMathOperator{\Suc}{Suc}

\newcommand{\cons}{\mathbin{\#}}

\begin{document}
\maketitle
\begin{abstract}
Ackermann's function can be expressed using an iterative algorithm, which essentially takes the form of a term rewriting system. Although the termination of this algorithm is far from obvious, its equivalence to the traditional recursive formulation---and therefore its totality---has a simple proof in Isabelle/HOL\@. This is a small example of formalising mathematics using a proof assistant, with a focus on the treatment of difficult recursions.

\smallskip
\textit{AMS classification numbers (MSC 2020)}: 03D99, 03B35, 68V20 

\end{abstract}

\section{Introduction}

The past few years have seen significant achievements in the mechanisation of mathematics~\cite{avigad-mechanization}, using proof assistants such as Coq and Lean. Here we  examine a simple example involving Ackermann's function: on how to prove the correctness of a system of rewrite rules for computing this function, using Isabelle.
The article also includes an introduction to the principles of implementing a proof assistant.

Formal models of computation include Turing machines, register machines and the general recursive functions. In such models, computations are reduced to basic operations such as writing symbols to a tape, testing for zero or adding or subtracting one. Because computations may terminate for some values and not others, partial functions play a major role and the domain of a partial function (i.e.\ the set of values for which the computation terminates) can be nontrivial~\cite{kleene52}. The primitive recursive functions---a subclass of the recursive functions---are always total.

In 1928, Wilhelm Ackermann exhibited a function that was obviously computable and total, yet could be proved not to belong to the class of primitive recursive functions~\cite[p.\ts272]{kleene52}. Simplified by R{\'o}zsa P\'eter and Raphael Robinson, it comes down to us in the following well-known form:
\begin{align*}
	A(0,n) & = n+1\\
	A(m+1,0) & = A(m,1)\\
	A(m+1,n+1) & = A(m,A(m+1,n)).
\end{align*}
In 1993, Szasz~\cite{szasz93} proved that Ackermann's function was not primitive recursive using a type theory based proof assistant called~ALF\@.

Isabelle/HOL~\cite{nipkow-concrete-semantics,isa-tutorial} is a proof assistant based on higher-order logic. Its underlying logic is much simpler than the type theories used in Coq for example. In particular, the notion of a recursive function is not primitive to higher-order logic but is derivable. We can introduce Ackermann's function to Isabelle/HOL as shown below. The specification invokes internal machinery to generate a low-level definition and derive the claimed identities from it. Here \isa{Suc} denotes the successor function for the natural numbers (type \isa{nat}).
\begin{isabelle}
\isacommand{fun}\ ack\ ::\ "[nat,nat]\ \isasymRightarrow \ nat"\ \isakeyword{where}\isanewline
\ \ "ack\ 0\ n\ \ \ \ \ \ \ \ \ \ \ \ \ =\ Suc\ n"\isanewline
|\ "ack\ (Suc\ m)\ 0\ \ \ \ \ \ \ =\ ack\ m\ 1"\isanewline
|\ "ack\ (Suc\ m)\ (Suc\ n)\ =\ ack\ m\ (ack\ (Suc\ m)\ n)"
\end{isabelle}
It is easy to see that the recursion is well defined and terminating. In every recursive call, either the first or the second argument decreases by one, suggesting a termination ordering: the lexicographic combination of $<$ (on the natural numbers) for the two arguments. 

Nevertheless, it's not straightforward to prove that Ackermann's function belongs to the class of computable functions in a formal sense.
Cutland~\cite[p.\ts46--7]{cutland-computability} devotes an entire page to the sketch of a construction to show that Ackermann's function could be computed using a register machine, before remarking that ``a sophisticated proof'' is available as an application of
  more advanced results, presumably the recursion theorem. 
This raises the question of whether Ackermann's function has some alternative definition that is easier to reason about, and in fact, iterative definitions exist. But then we must prove that the recursive and iterative definitions are equivalent.
  
The proof is done using the function definition facilities of Isabelle/HOL and is a good demonstration of their capabilities to the uninitiated. But first, we need to consider how function definitions are handled in Isabelle/HOL and how the later relates to symbolic logic.

\section{Recursive function definitions in Isabelle/HOL}

Isabelle's higher-order logic is a form of Church's simple type theory~\cite{church40}. As with Church, it is based on the typed $\lambda$-calculus with function types (written $\alpha\to\beta$: Greek letters range over types) and a type of booleans (written \isa{bool}). 
Again following Church, the axiom of choice is provided through Hilbert's epsilon operator $\epsilon x. \phi$, denoting some~$a$ such that $\phi(a)$ if such exists and otherwise any value. 

For Church, all types were built up from the booleans and a type of individuals, keeping types to the minimum required for consistency.  Isabelle/HOL has a multiplicity of types in the spirit of functional programming, with numeric types \isa{nat}, \isa{int}, \isa{real}, among countless others.
Predicates have types of the form $\alpha\to\isa{bool}$, but for reasons connected with performance, the distinct but equivalent type $\alpha\,\isa{set}$ is provided for sets of elements of type~$\alpha$.

Gordon~\cite{mgordon86} pioneered the use of simple type theory for verifying hardware.
His first computer implementation, and the later HOL Light~\cite{harrison-hol-light}, hardly deviate from Church. Constants can be introduced, but they are essentially abbreviations. The principles for defining new types do not stretch things much further: they allow the declaration of a new type corresponding to what Church would have called ``a non-empty class given by a propositional function'' (a predicate over an existing type). These principles, some criticisms of them and proposed alternatives are explored by Arthan~\cite{arthan-definitions}.

The idea of derivations schematic over types is already implicit in Church (``typical ambiguity''), and in most implementations is placed on a formal basis by including type variables in the calculus.
Then all constructions involving types can be schematic, or \textit{polymorphic}, allowing for example a family of types of the form $\alpha\,\isa{list}$, conventionally written in postfix notation. 
Refining the notion of polymorphism to allow classes of type variables associated with axioms---so-called \textit{axiomatic type classes}---is a major extension to Church's original conception, and has required a thoroughgoing analysis~\cite{kuncar-consistent-foundation-jar}. However, those extensions are not relevant here, where we are only interested in finite sequences of integers.

There are a number of ways to realise a logical calculus on a computer. At one extreme, the implementer might choose a fast, unsafe language such as C and write arbitrarily complex code, implementing algorithms that have been shown to be sound with respect to the chosen calculus. Automatic theorem provers follow this approach.
Most proof assistants, including Isabelle, take the opposite extreme and prioritise correctness. 
The implementer codes the axioms and inference rules of the calculus in something approaching their literal form:
providing syntactic operations on types and terms while encapsulating
the logical rules within a small, dedicated \textit{proof kernel}.
This \textit{LCF architecture}~\cite{gordon-tactics-milner} requires a safe programming language so that the proof kernel---which has the exclusive right to declare a formula to be a theorem---can be protected from any bugs in the rest of the system.

Formal proofs are frequently colossal, so most proof assistants provide automation.
In Isabelle, the \isa{auto} proof method simplifies arithmetic expressions, expands functions when they are applied to suitable arguments and performs simple logical reasoning.
Users can add automation to Isabelle by writing code for say a decision procedure, but such code (like \isa{auto} itself) must lie outside the proof kernel and must reduce its proofs to basic inferences so that they can pass through the kernel. In this way, the LCF architecture eliminates the need to store the low-level proofs themselves, a vital space saving even in the era of 32 GB laptops.

 Sophisticated principles for defining inductive sets, recursive functions with pattern matching and recursive types can be reduced to pure higher-order logic.
 In accordance with the LCF architecture, such definitions are translated into the necessary low-level form by Isabelle/HOL code that lies outside the proof kernel.  This code defines basic constructions, from which it then proves desired facts, such as the function's recursion equations.
 
 In mathematics, a recursive function must always be shown to be well defined.
 Non-terminating recursion equations cannot be asserted unconditionally, since they could yield a contradiction: consider $f(m,n) = f(n,m)+1$, which implies $f(0,0) = f(0,0)+1$. 
 Isabelle/HOL's function package, due to Alexander Krauss~\cite{krauss-partial-recursive}, reduces recursive function definitions to inductively defined relations.
A recursive function~$f$ is typically partial, so the package also defines its \textit{domain}~$D_f$, the set of values for which $f$ obeys its recursion equations.%
\footnote{Since there are no partial functions in higher-order logic, $f(x)$ yields an arbitrary value if $x\not\in D_f$.}

The idea of inductive definitions should be familiar, as when we say the set of theorems is inductively generated by the given axioms and inference rules.
 Formally, a set $I(\Phi)$ is \textit{inductively defined} with respect to a collection~$\Phi$ of rules provided it is closed under $\Phi$ and is the least such set~\cite{aczel77}. 
In higher-order logic, $I(\Phi)$ can be defined as the intersection of all sets closed under a collection of rules: $I(\Phi) = \bigcap \{ A \mid A \text{ is $\Phi$-closed}\}$. The minimality of $I(\Phi)$, namely that $I(\Phi)\subseteq A$ if $A$ is $\Phi$-closed, gives rise to a familiar principle for proof by induction. Even Church~\cite{church40} included a construction of the natural numbers.
Isabelle provides a package to automate inductive definitions~\cite{paulson-fixedpt-milner}.
  
Krauss' function package~\cite{krauss-partial-recursive} includes many refinements so as to handle straightforward function definitions---like the one shown in the introduction---without fuss. Definitions go through several stages of processing. The specification of a function~$f$ is examined, following the recursive calls, to yield inductive definitions of its graph $G_f$ and domain~$D_f$. The package proves that $G_f$ corresponds to a well-defined function on its domain. It is then possible to define $f$ formally in terms of~$G_f$ and to derive the desired recursion equations, each conditional on the function being applied within its domain. The refinements alluded to above include dealing with pattern matching and handling easy cases of termination, where the domain can be hidden. But in the example considered below, we are forced to prove termination ourselves through a series of inductions. 

For a simple example~\cite[\S3.5.4]{krauss-partial-recursive}, consider the everywhere undefined function given by $U(x) = U(x)+1$. The graph is defined inductively by the rule
\[ (x, h(x)) \in G_U \Longrightarrow (x, h(x)+1) \in G_U.\]
Similarly, the domain is defined inductively by the rule
\[ x \in D_U \Longrightarrow x \in D_U.\]
It should be obvious that $G_U$ and $D_U$ are both empty and that the evaluation rule $x \in D_U \Longrightarrow U(x) = U(x)+1$ holds vacuously. But we can also see how less trivial examples might be handled, as in the extended example that follows.
 
\section{An Iterative Version of Ackermann's Function}

A \textit{list} is a possibly empty finite sequence, written $[x_1,\ldots,x_n]$ or equivalently $x_1\cons \cdots\cons x_n\cons []$.
Note that $\cons$ is the operation that extends a list from the front with a new element.
We can write an iterative definition of~$A$ in terms of the following recursion on lists:
\begin{align*}
	n\cons 0\cons L &\longrightarrow \Suc n \cons  L\\
	0\cons \Suc m\cons L &\longrightarrow 1\cons  m \cons L\\
	\Suc n\cons \Suc m\cons L &\longrightarrow n\cons \Suc m\cons  m \cons L
\end{align*}
the idea being to replace the recursive calls by a stack. We intend that a computation starting with a two-element list will yield the corresponding value of Ackermann's function:
\[ [n,m] \longrightarrow^* [A(m,n)]. \]
An execution trace for $A(2,3)$ looks like this:
\begin{quote}\footnotesize
3 2\\
2 2 1\\
1 2 1 1\\
0 2 1 1 1\\
1 1 1 1 1\\
0 1 0 1 1 1\\
1 0 0 1 1 1\\
2 0 1 1 1\\
3 1 1 1\\
2 1 0 1 1\\
1 1 0 0 1 1\\
0 1 0 0 0 1 1\\
1 0 0 0 0 1 1\\
2 0 0 0 1 1\\
3 0 0 1 1\\
4 0 1 1\\
5 1 1\\
4 1 0 1\\
3 1 0 0 1\\
2 1 0 0 0 1\\
1 1 0 0 0 0 1\\
0 1 0 0 0 0 0 1\\
1 0 0 0 0 0 0 1\\
2 0 0 0 0 0 1\\
3 0 0 0 0 1\\
4 0 0 0 1\\
5 0 0 1\\
6 0 1\\
7 1\\
6 1 0\\
5 1 0 0\\
4 1 0 0 0\\
3 1 0 0 0 0\\
2 1 0 0 0 0 0\\
1 1 0 0 0 0 0 0\\
0 1 0 0 0 0 0 0 0\\
1 0 0 0 0 0 0 0 0\\
2 0 0 0 0 0 0 0\\
3 0 0 0 0 0 0\\
4 0 0 0 0 0\\
5 0 0 0 0\\
6 0 0 0\\
7 0 0\\
8 0\\
9
\end{quote}

We can regard these three reductions as constituting a term rewriting system~\cite{baader-nipkow}, subject to the proviso that they can only rewrite at the front of the list. Equivalently, each rewrite rule can be imagined as beginning with an anchor symbol, say~$\Box$: 
\begin{align*}
	\Box \cons n\cons 0\cons L &\longrightarrow \Box \cons \Suc n \cons  L\\
	\Box \cons 0\cons \Suc m\cons L &\longrightarrow \Box \cons 1\cons  m \cons L\\
	\Box \cons \Suc n\cons \Suc m\cons L &\longrightarrow \Box \cons n\cons \Suc m\cons  m \cons L
\end{align*}

A term rewriting system is a model of computation in itself.
But termination isn't obvious here. In the first rewrite rule above, the head of the list gets bigger
 while the list gets shorter, suggesting that the length of the list should be the primary
termination criterion. But in the third rewrite rule, the list gets longer. One might imagine a more sophisticated approach to termination based on multisets or ordinals; these however could lead nowhere because the second rewrite allows $0\cons 1\cons L \longrightarrow 1\cons 0 \cons L$ and often these approaches ignore the order of the list elements.

Although some natural termination ordering might be imagined to exist,%
\footnote{Ren\'e Thiemann has kindly run some tests using rewrite system termination checkers. Without the anchors, the rewrite system is non-terminating because rewrite rules can be applied within a list. With the anchors, no currently existing termination checker reaches a conclusion.}
this system is an excellent way to demonstrate another approach to proving termination: by explicit reasoning about the domain of definition. It is easy, using Isabelle/HOL's function definition package~\cite{krauss-partial-recursive}.

\section{The Iterative Version in Isabelle/HOL}

We would like to formalise the iterative computation described above as a recursive function, but we don't know that it terminates. 
Isabelle allows the following form, with the keyword \isa{domintros}, indicating that we wish to \textit{defer} the termination proof and reason explicitly about the function's domain.  Our goal is to show that the set is \textit{universal} (for its type).
\begin{isabelle}
\isacommand{function}\ (domintros)\ ackloop\ ::\ "nat\ list\ \isasymRightarrow \ nat"\ \isakeyword{where}\isanewline
\ \ "ackloop\ (n\ \#\ 0\ \#\ L)\ \ \ \ \ \ \ \ \ =\ ackloop\ (Suc\ n\ \#\ L)"\isanewline
|\ "ackloop\ (0\ \#\ Suc\ m\ \#\ L)\ \ \ \ \ =\ ackloop\ (1\ \#\ m\ \#\ L)"\isanewline
|\ "ackloop\ (Suc\ n\ \#\ Suc\ m\ \#\ L)\ =\ ackloop\ (n\ \#\ Suc\ m\ \#\ m\ \#\ L)"\isanewline
|\ "ackloop\ [m]\ =\ m"\isanewline
|\ "ackloop\ []\ =\ \ 0"
\end{isabelle}

The domain, which is called \isa{ackloop\_dom}, is generated according to the recursive calls. It is defined inductively to satisfy the following properties:%
\footnote{For clarity, \isa{Suc\ 0} has been replaced by~\isa{1}.}
\begin{isabelle}
ackloop\_dom\ (Suc\ n\ \#\ L)\ \isasymLongrightarrow \ ackloop\_dom\ (n\ \#\ 0\ \#\ L)\isasep\isanewline%
ackloop\_dom\ (1\ \#\ m\ \#\ L)\ \isasymLongrightarrow \ ackloop\_dom\ (0\ \#\ Suc\ m\ \#\ L)\isasep\isanewline%
ackloop\_dom\ (n\ \#\ Suc\ m\ \#\ m\ \#\ L)\ \isasymLongrightarrow \ ackloop\_dom\ (Suc\ n\ \#\ Suc\ m\ \#\ L)\isasep\isanewline%
ackloop\_dom\ [m]\isasep\isanewline%
ackloop\_dom\ []
\end{isabelle}
For example, the first line states that if \isa{ackloop} terminates for \isa{Suc\ n\ \#\ L} then it will also terminate for \isa{n\ \#\ 0\ \#\ L}, as we can see for ourselves by looking at the first line of \isa{ackloop}. The second and third lines similarly follow the recursion. The last two lines are unconditional because there is no recursion.

It's obvious that \isa{ackloop\_dom} holds for all lists shorter than two elements. Its properties surely allow us to prove instances for longer lists (thereby establishing termination of \isa{ackloop} for those lists), but how? At closer examination, remembering that \isa{ackloop} represents the recursion of Ackermann's function, we might come up with the following lemma:
\begin{isabelle}
\ \ ackloop\_dom\ (ack\ m\ n\ \#\ L)\ \isasymLongrightarrow \ ackloop\_dom\ (n\ \#\ m\ \#\ L)
\end{isabelle}
This could be the solution, since it implies that \isa{ackloop} terminates on the list $n\cons m\cons L$ provided it terminates on $A(m,n)\cons L$, a shorter list. And indeed it can easily be proved by mathematical induction on~$m$ followed by a further induction on~$n$. If $m=0$ then it simplifies to the first \isa{ackloop\_dom} property:
\begin{isabelle}
\ \ ackloop\_dom\ (Suc n\ \#\ L)\ \isasymLongrightarrow \ ackloop\_dom\ (n\ \#\ 0\ \#\ L)
\end{isabelle}
In the $\Suc m$ case, after the induction on~$n$, the $n=0$ case simplifies to
\begin{isabelle}
\ \ ackloop\_dom\ (ack\ m\ 1\ \#\ L)\ \isasymLongrightarrow \ ackloop\_dom\ (0\ \#\ Suc\ m\ \#\ L)
\end{isabelle}
but from \isa{ackloop\_dom\ (ack\ m\ 1\ \#\ L)} the induction hypothesis yields
\isa{ackloop\_dom\ (1\ \#\ m\ \#\ L)}, from which we obtain \isa{ackloop\_dom\ (0\ \#\ Suc\ m\ \#\ L)} by the second \isa{ackloop\_dom} property. The $\Suc n$ case is also straightforward:
\begin{isabelle}
\ ackloop\_dom\ (ack\ (Suc\ m)\ (Suc\ n)\ \#\ L) \isasymLongrightarrow \ ackloop\_dom\ (Suc\ n\ \#\ Suc\ m\ \#\ L)
\end{isabelle}
It needs the third \isa{ackloop\_dom} property and both induction hypotheses. The details are left as an exercise.

In Isabelle, the lemma proved above can be proved in one line, thanks to a special induction rule: \isa{ack.induct}. The definition of a function~$f$ in Isabelle automatically yields an induction rule customised to the recursive calls, derived from the inductive definition of~$G_f$. For \isa{ack}, it allows us to prove any formula $P(x,y)$ from the three premises 
\begin{align*}
& \forall n\,P(0,n)\\	
& \forall m\, [P(m,1) \Longrightarrow P (m+1, 0)]\\
& \forall m\,n\, [P (m+1, n) \land P(m, A(m+1,n)) \Longrightarrow P (m+1,n+1)]
\end{align*}
Using this induction rule, our lemma follows immediately by simple rewriting:
\begin{isabelle}
\isacommand{lemma}\ ackloop\_dom\_longer:\isanewline
\ \ "ackloop\_dom\ (ack\ m\ n\ \#\ L)\ \isasymLongrightarrow \ ackloop\_dom\ (n\ \#\ m\ \#\ L)"\isanewline
\ \ \isacommand{by}\ (induction\ m\ n\ arbitrary:\ L\ rule:\ ack.induct)\ auto
\end{isabelle}
Let's examine this proof. In the induction, $P(m,n)$ is the formula
\begin{quote}
\begin{isabelle}
$\forall$L [ ackloop\_dom\ (ack\ m\ n\ \#\ L)\ \isasymLongrightarrow \ ackloop\_dom\ (n\ \#\ m\ \#\ L) ]
\end{isabelle}	
\end{quote}
In most difficult case, $P (m+1,n+1)$, the left-hand side is
\begin{quote}
\begin{isabelle}
ackloop\_dom (ack (Suc m) (Suc n) \# L)\\
\textrm{\normalfont $\to$ (by evaluation)}\\
ackloop\_dom (ack m (ack (Suc m) n) \# L)\\ 
\textrm{\normalfont $\to$ (second induction hypothesis)}\\
ackloop\_dom (ack (Suc m) n \# m \# L)\\  
\textrm{\normalfont $\to$ (first induction hypothesis)}\\
ackloop\_dom (n \# Suc m \# m \# L)\\     
\textrm{\normalfont $\to$ (definition of \isa{ackloop\_dom})}\\
	ackloop\_dom (Suc n \# Suc m \# L)
\end{isabelle}	
\end{quote}
And this is the right-hand side of $P (m+1,n+1)$. 

It must be stressed that when typing in the Isabelle proof shown above for lemma \isa{ackloop\_dom\_longer}, I did not have this or any derivation in mind. Experienced users know that properties of a recursive function~\isa{f} often have extremely simple proofs by induction on~\isa{f.induct} followed by \isa{auto} (basic automation), so they type the corresponding Isabelle commands without thinking. We are gradually managing to shift the burden of thinking to the computer.

\section{Completing the Proof}

Given the lemma just proved, it's clear that every list~$L$ satisfies \isa{ackloop\_dom} by induction on the length~$l$ of~$L$: if $l<2$ then the result is immediate, and otherwise it has the form $n\cons m \cons L'$, which the lemma reduces to $A(m,n) \cons L'$ and we are finished by the induction hypothesis.

A slicker proof turns out to be possible. Consider what \isa{ackloop} is actually designed to do: to replace the first two list elements, $n$ and $m$, by $A(m,n)$.
The following function codifies this point.
\begin{isabelle}
\isacommand{fun}\ acklist\ ::\ "nat\ list\ \isasymRightarrow \ nat"\ \isakeyword{where}\isanewline
\ \ "acklist\ (n\#m\#L)\ =\ acklist\ (ack\ m\ n\ \#\ L)"\isanewline
|\ "acklist\ [m]\ =\ m"\isanewline
|\ "acklist\ []\ =\ \ 0"
\end{isabelle}

As mentioned above, recursive function definitions automatically provide us with a customised induction rule. In the case of \isa{acklist}, it performs exactly the case analysis sketched at the top of this section. So this proof is also a single induction followed by automation. Note the reference to \isa{ackloop\_dom\_longer}, the lemma proved above.
\begin{isabelle}
\isacommand{lemma}\ ackloop\_dom:\ "ackloop\_dom\ L"\isanewline
\ \ \isacommand{by}\ (induction\ L\ rule:\ acklist.induct)\ (auto\ simp:\ ackloop\_dom\_longer)
\end{isabelle}
It is possible to reconstruct the details of this proof by running it interactively, as was done in the previous section. But perhaps it is better to repeat that these Isabelle commands were typed without having any detailed proof in mind but simply with the knowledge that they were likely to be successful.

Now that \isa{ackloop\_dom} is known to hold for arbitrary~\isa{L}, we can issue a command to inform Isabelle that \isa{ackloop} is a total function satisfying \textit{unconditional} recursion equations. We mention the termination result just proved.
 \begin{isabelle}
\isacommand{termination}\ ackloop\isanewline
\ \ \isacommand{by}\ (simp\ add:\ ackloop\_dom)
\end{isabelle}

The equivalence between \isa{ackloop} and \isa{acklist} is another one-line induction proof. The  induction rule for \isa{ackloop} considers the five cases of that function's definition, which---as we have seen twice before---are all proved automatically.
\begin{isabelle}
\isacommand{lemma}\ ackloop\_acklist:\ "ackloop\ L\ =\ acklist\ L"\isanewline
\ \ \isacommand{by}\ (induction\ L\ rule:\ ackloop.induct)\ auto
\end{isabelle}
The equivalence between the iterative and recursive definitions of Ackermann's function is now immediate.
\begin{isabelle}
\isacommand{theorem}\ ack:\ "ack\ m\ n\ =\ ackloop\ [n,m]"\isanewline
\ \ \isacommand{by}\ (simp\ add:\ ackloop\_acklist)
\end{isabelle}

We had a function that obviously terminated but was not obviously computable (in the sense of Turing machines and similar formal models) and another function that was obviously computable but not obviously terminating. The proof of the termination of the latter has led immediately to a proof of equivalence with the former.

Anybody who has used a proof assistant knows that machine proofs are generally many times longer than typical mathematical exposition. Our example here is a rare exception.

\paragraph*{Acknowledgements.}
This work was supported by the ERC Advanced Grant ALEXANDRIA (Project GA 742178). Ren\'e Thiemann investigated the termination of the corresponding rewrite systems. I am grateful to the editor, Peter Dybjer, and the referee for their comments.


\bibliographystyle{plain}
\bibliography{string,atp,general,isabelle,theory,crossref}

\begin{thebibliography}{10}

\bibitem{aczel77}
Peter Aczel.
\newblock An introduction to inductive definitions.
\newblock In J.~Barwise, editor, {\em Handbook of Mathematical Logic}, pages
  739--782. North-Holland, 1977.

\bibitem{arthan-definitions}
Rob Arthan.
\newblock On definitions of constants and types in {HOL}.
\newblock {\em Journal of Automated Reasoning}, 56(3):205--219, March 2016.

\bibitem{avigad-mechanization}
Jeremy Avigad.
\newblock Opinion: The mechanization of mathematics.
\newblock {\em Notices of the American Mathematical Society}, 65(6):681--690,
  2018.
\newblock Online at
  \url{http://www.ams.org/journals/notices/201806/rnoti-p681.pdf}.

\bibitem{baader-nipkow}
Franz Baader and Tobias Nipkow.
\newblock {\em Term Rewriting and All That}.
\newblock Cambridge University Press, 1998.

\bibitem{church40}
Alonzo Church.
\newblock A formulation of the simple theory of types.
\newblock {\em Journal of Symbolic Logic}, 5:56--68, 1940.

\bibitem{cutland-computability}
Nigel Cutland.
\newblock {\em Computability: An Introduction to Recursive Function Theory}.
\newblock Cambridge University Press, 1980.

\bibitem{gordon-tactics-milner}
M.~J.~C. Gordon.
\newblock Tactics for mechanized reasoning: A commentary on {Milner} (1984)
  {\textquoteleft}{The} use of machines to assist in rigorous
  proof{\textquoteright}.
\newblock {\em Philosophical Transactions of the Royal Society of London A:
  Mathematical, Physical and Engineering Sciences}, 373(2039), 2015.

\bibitem{mgordon86}
Michael J.~C. Gordon.
\newblock Why higher-order logic is a good formalism for specifying and
  verifying hardware.
\newblock In G.~Milne and P.~A. Subrahmanyam, editors, {\em Formal Aspects of
  {VLSI} Design}, pages 153--177. North-Holland, 1986.

\bibitem{harrison-hol-light}
John Harrison.
\newblock {HOL Light}: An overview.
\newblock In Stefan Berghofer, Tobias Nipkow, Christian Urban, and Makarius
  Wenzel, editors, {\em Theorem Proving in Higher Order Logics}, pages 60--66.
  Springer, 2009.

\bibitem{kleene52}
S.~C. Kleene.
\newblock {\em Introduction to Metamathematics}.
\newblock North-Holland, 1952.

\bibitem{krauss-partial-recursive}
Alexander Krauss.
\newblock Partial and nested recursive function definitions in higher-order
  logic.
\newblock {\em Journal of Automated Reasoning}, 44(4):303--336, 2010.

\bibitem{kuncar-consistent-foundation-jar}
Ond{\v r}ej Kun{\v c}ar and Andrei Popescu.
\newblock A consistent foundation for {Isabelle/HOL}.
\newblock {\em Journal of Automated Reasoning}, 62(4):531--555, 2019.

\bibitem{nipkow-concrete-semantics}
Tobias Nipkow and Gerwin Klein.
\newblock {\em Concrete Semantics with {Isabelle/HOL}}.
\newblock Springer, 2014.

\bibitem{isa-tutorial}
Tobias Nipkow, Lawrence~C. Paulson, and Markus Wenzel.
\newblock {\em Isabelle/HOL: A Proof Assistant for Higher-Order Logic}.
\newblock Springer, 2002.
\newblock Online at
  \url{http://isabelle.in.tum.de/dist/Isabelle/doc/tutorial.pdf}.

\bibitem{paulson-fixedpt-milner}
Lawrence~C. Paulson.
\newblock A fixedpoint approach to (co)inductive and (co)datatype definitions.
\newblock In Gordon Plotkin, Colin Stirling, and Mads Tofte, editors, {\em
  Proof, Language, and Interaction: Essays in Honor of {Robin Milner}}, pages
  187--211. MIT Press, 2000.

\bibitem{szasz93}
Nora Szasz.
\newblock A machine checked proof that {Ackermann's} function is not primitive
  recursive.
\newblock In {G\'erard} Huet and Gordon Plotkin, editors, {\em Logical
  Environments}, pages 317--338. Cambridge University Press, 1993.

\end{thebibliography}

\end{document}